\documentstyle[preprint,aps,epsf]{revtex}

\begin{document} \author{T. Jacobs, Balam A. Willemsen
and S. Sridhar} \address{Physics Department, Northeastern University,
Boston, MA 02115, \\ } \author{R. Nagarajan, L. C. Gupta, Z. Hossain
and Chandan Mazumdar} \address{Tata Inst. of Fundamental Research,
Colaba, Bombay, India, \\ } \author{P. C. Canfield and B. K. Cho}
\address{Ames Laboratory and Dept. of Physics and Astronomy, Iowa
State \\University, Ames, IA.}  \title{Microwave Properties of
Borocarbide Superconductors LnNi$_2$B$_2$C \\ (Ln = Y, Er, Tm, Ho)}
\date{\today } \maketitle

\begin{abstract} We report measurements of the microwave surface
impedance of the borocarbide family of superconductors LnNi$_2$B$_2$C
(Ln$=$Y, Er, Tm, Ho). The experiments enable direct measurements of
the superfluid density, and are particularly sensitive to the
influence of magnetic pairbreaking. In HoNi$_2$B$_2$C the
antiferromagnetic transition is clearly observed at zero field, and
leads to a drastic reduction of the superfluid density, which recovers
at lower temperatures.
 In ErNi$_2$B$_2$C the antiferromagnetic transition is not seen in
zero field data. Magnetic effects are responsible for anomalies in the
low temperature surface impedance below approximately 4K in
HoNi$_2$B$_2$C and TmNi$_2$B$_2$C. The temperature dependence of the
microwave impedance disagrees with simple BCS calculations.
\end{abstract}

Recently superconductivity was discovered in quaternary multiphase
Y-Ni-B-C \cite{RNagarajan94}, in multiphase Y-Pd-B-C system
\cite{RJCava94} and in single phase LnNi$_2$B$_2$C (Ln=Y,Ho,Er,Tm,Lu)
\cite{RJCava94b}. These materials are a class of intermetallic
superconductors. While some members of the family, e.g. YNi$_2$B$_2$C,
appear to show conventional superconducting behavior
\cite{SACarter94,pcc5}, other members, with Ln$=$Tm, Er, Ho, undergo
an antiferromagnetic (AFM) transition at $T_N$, below the
superconducting transition at $T_c$. Thus in addition to the important
question of the nature of superconductivity issues concerning the
interplay of magnetism and superconductivity arise in these materials.

Microwave measurements of superconductors yield unique information
regarding the superfluid density, the quasiparticles and the nature of
the pairing \cite{Sridhar89}. Despite the fact that magnetic
superconductors such as ErRhB$_4$ have been known for about two
decades, there have been few microwave studies of such
superconductors. In this paper we report the first measurements of the
microwave response of several members of the family,
LnNi$_2$B$_2$Cwith $Ln=Y,\,Er,\,Tm,\,Ho$.

Single crystals of LnNi$_2$B$_2$C were grown out of Ni$_2$B
flux. Details are provided in Ref.~\cite{pcc5}. The crystals have a
plate-like morphology and X-ray diffraction indicates that they have
their crystallographic c-axis perpendicular to the surface of the
plate. Powder X-ray diffraction on ground single crystals indicate
that there is a small amount of Ni$_2$B flux contaminating the
crystal, probably on the surface. This contamination is estimated to
be below the 5\% level. These crystals have been extensively
characterized by a variety of techniques \cite{pcc5,pcc6,pcc7,pcc8}
besides the microwave studies discussed here. Polycrystalline samples
of YNi$_2$B$_2$C were also prepared and measured.

Microwave measurements were carried out in a Nb cavity using a high
precision ``hot finger'' method \cite{Sridhar88}, in which the cavity
is maintained at 2$\,$K while the sample temperature is varied from
2$\,$K to as high as 200$\,$K. The surface resistance of the sample is
obtained from measurements of the resonator Q using $R_s=\Gamma
\,[Q_s^{-1}(T)\,-\,Q_0^{-1}(T)]$, where $Q_s$ and $Q_o$ are
respectively the resonator Q's with and without the sample. Similarly
changes in the reactance are obtained from $\Delta X_s=\mu_0\omega
\Delta \lambda = (-2\Gamma /f_0)\,[f_s(T)-f_0(T)]$, where $f_s$ and
$f_0$ represent the resonant frequency with and without the sample. To
get the absolute value of the reactance, an indeterminate constant
$X_0$ needs to be added.  In materials which obey the skin depth limit
in the normal state, one can use the criterion that $R_n=X_n$ above
$T_c$ to determine $X_0$ and so the absolute value of $X_s$.  The
geometric constant $\Gamma $ was calculated, and confirmed using
measurements on known samples of Cu. The method has been extensively
validated via measurements on a variety of samples, from conventional
low temperature superconductors, to high temperature superconducting
crystals and films \cite{SOxx94}.

Because we are able to measure both the real and imaginary parts of
the impedance $Z_s=R_s+\mathrm{i}X_s$, it is possible to obtain the
complex conductivity $\sigma _s=\sigma _1-\mathrm{i}\sigma _2$ from
the present measurements, using the relation
$Z_s=R_n\sqrt{2\mathrm{i}/(\sigma _s/\sigma _n)}$, where $R_n$ and
$\sigma _n$ are the normal state surface resistance and conductivity
respectively. Of particular importance is the imaginary part $\sigma
_2$. This quantity is a measure of the superfluid density $n_s$ which
is proportional to $\sigma _2$.  It is also related to the penetration
depth via $\sigma _2=1/\mu _0\omega \lambda^2$.

\begin{figure} \epsfxsize=8.5cm \epsffile{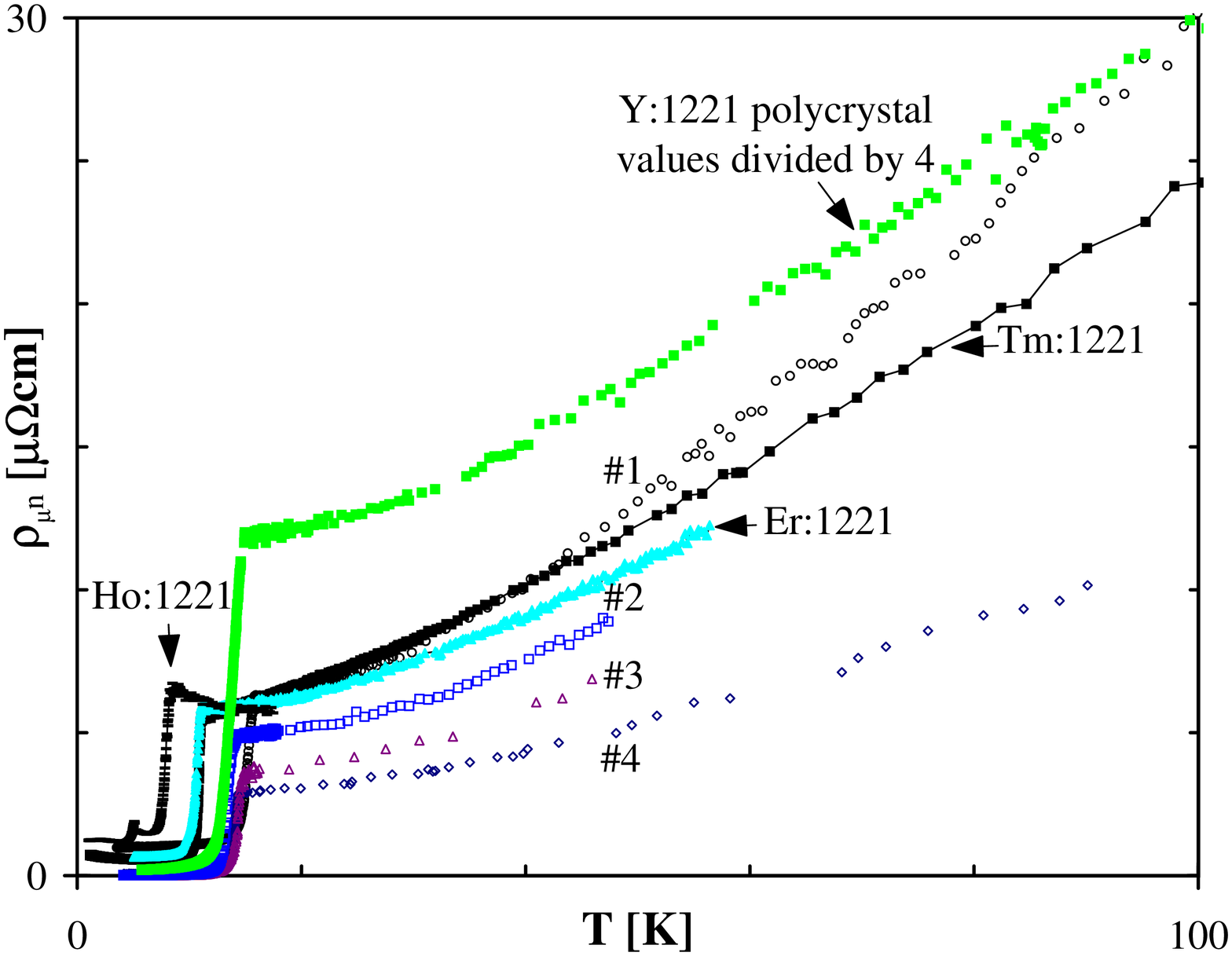}
\caption{Microwave resistivity of LnNi$_2$B$_2$ (Ln$ = $Y, Er, Tm, Ho)
in the normal and superconducting states extracted from the surface
resistance. $\#$1 to $\#$4 are different YNi$_2$B$_2$C single crystal
samples.}  \label{Fig1} \end{figure}

The microwave resistivity $\rho _{\mu n}(T)$ in the normal state, see
Fig.~\ref{Fig1}, was obtained from the surface resistance using
$R_n=\sqrt{ \pi \mu _0f\rho _{\mu n}}$ where $\mu _0$ is the
permeability of free space.  The data is in good agreement with dc
measurements \cite{DNaugle95}, confirming that the classical
skin-depth regime applies in the normal state.

{\bf YNi$_2$B$_2$C}

\begin{figure} \epsfxsize=8.5cm \epsffile{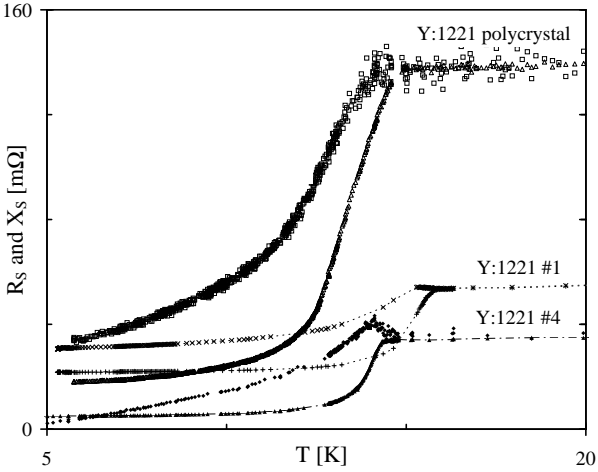} \caption{$R_s$ and
$X_s$ of YNi$_2$B$_2$C in the superconducting state for a polycrystal
and single crystals (\#1 and \#4).}  \label{Fig2} \end{figure}

 The superconducting surface resistance $R_s(T)$ of several samples of
YNi$_2$B$_2$C are shown in Fig.~\ref{Fig2}. The polycrystal shows
higher $R_n$ and also a slightly broader transition than single
crystals.
 It is evident
that $R_s$ does not go to zero as $T\rightarrow 0$. Thus there is a
residual surface resistance which is of the order of $10-20\,m\Omega $
and is, possibly, extrinsic in origin. This residual resistance does
depend on sample characteristics such as crystallinity and surface
preparation, being lowest for the polished single crystal $\#4$. This
is probably due to slight flux contamination which leads to a
temperature-independent microwave loss.

Fig.~\ref{Fig2} also shows the surface reactance $X_s(T)$. The
absolute value was obtained by normalizing to $X_n=R_n$ above $T_c$. A
striking feature of the reactance $X_s$ data is the peak near
$T_c$. This can be understood from simple models of the
superconducting state such as two- fluid or BCS (see below). This is
in fact occasionally observed in other superconductors, and is due to
buildup of superfluid, which leads to a situation where $\lambda
>\delta _n$, the superconducting penetration depth is bigger than the
normal state skin depth. This results in a peak in $X_s$ which we have
observed in Y$_1$Ba$_2$Cu$_3$O$_{1-\delta}$ \cite{Chen94} also. If we
substract the residual resistance from the low temperature reactance
we can estimate a zero-temperature penetration depth
$\lambda(0)=1100\AA$ for sample $\#1$.  Although this method is not
very sensitive to the absolute magnitudes the agreement with other
estimates of $\lambda (0)=1200\AA $ \cite {pcc5} is good.

 {\bf TmNi$_2$B$_2$C}

\begin{figure} \epsfxsize=8.5cm \epsffile{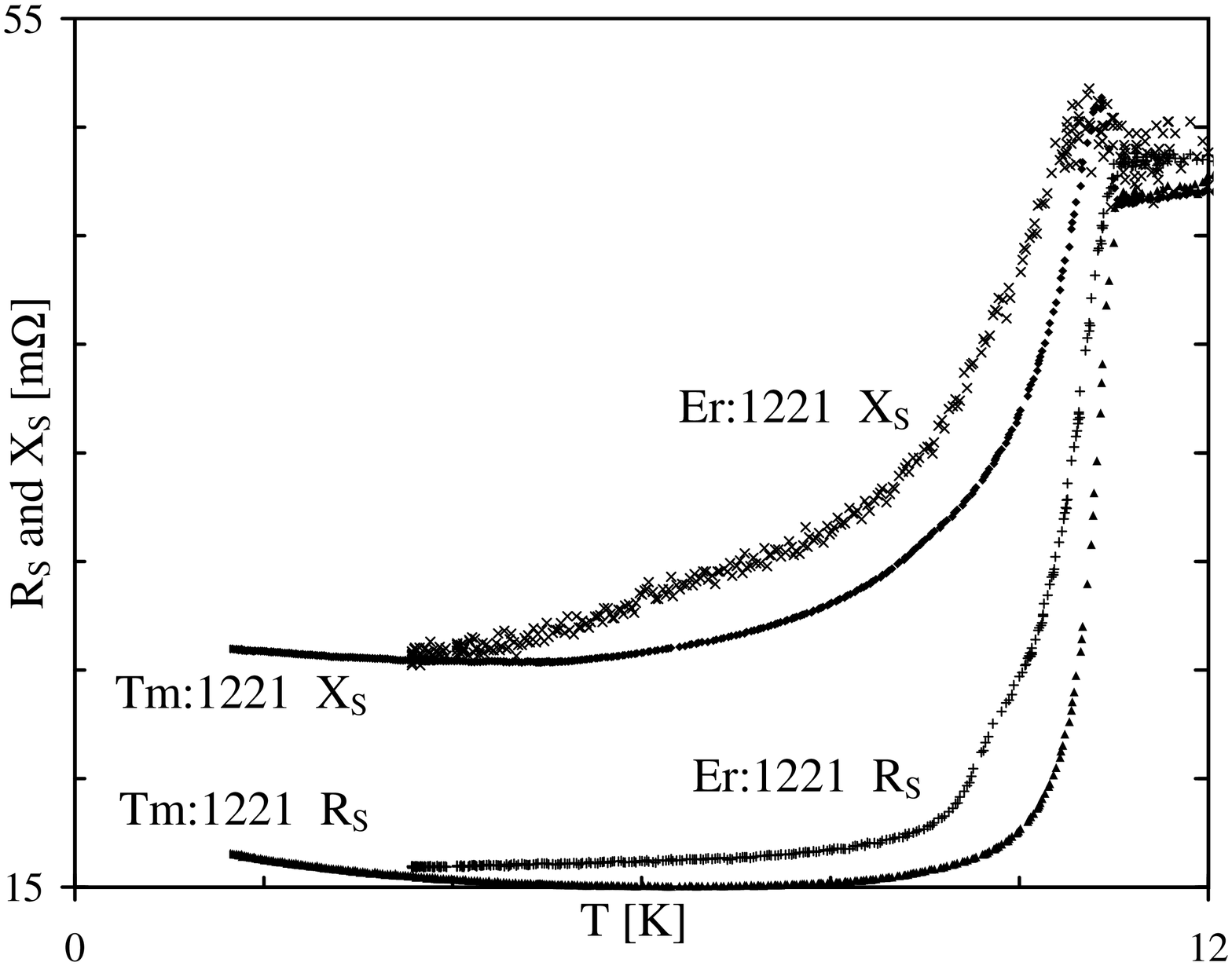} \caption{$R_s$
and $X_s$ of TmNi$_2$B$_2$C and ErNi$_2$B$_2$C.}  \label{Fig3}
\end{figure}

 While the overall behavior of TmNi$_2$B$_2$C near and slightly below
$T_c$ is similar to the $Y$ compound (see Fig.~\ref{Fig3}), a
noteworthy feature is the {\em increase} of $R_s$ at low
temperatures. This is due to increasing magnetic coherence which
results in a AFM transition at $1.5K$ \cite{pcc1}, which is however
just below the lower limit of our apparatus.  This is also consistent
with a decrease of $H_{c2}$ observed in the same temperature region
\cite{HEisaki94}. The data demonstrate that the microwave measurements
can provide a measure of the magnetic scattering through the influence
on the surface impedance of the superconducting state.

 {\bf ErNi$_2$B$_2$C}

 The response of this compound appears similar to that of the Y
compound as shown in Fig.~\ref{Fig3}. Interestingly the AFM transition
which should occur at $6.0K$ \cite{BKCho95,HEisaki94} as observed in
specific heat, dc-resistivity and magnetic susceptibility
measurements, does not appear in the microwave data. Thus in this
compound, the AFM transition is not accompanied by pairbreaking in
zero field.  There is however a weak shoulder in the $R_s$ data at
around $9.5\,$K which is not associated with a feature in any other
measurement.

 {\bf HoNi$_2$B$_2$C}

\begin{figure} \epsfxsize=8.5cm \epsffile{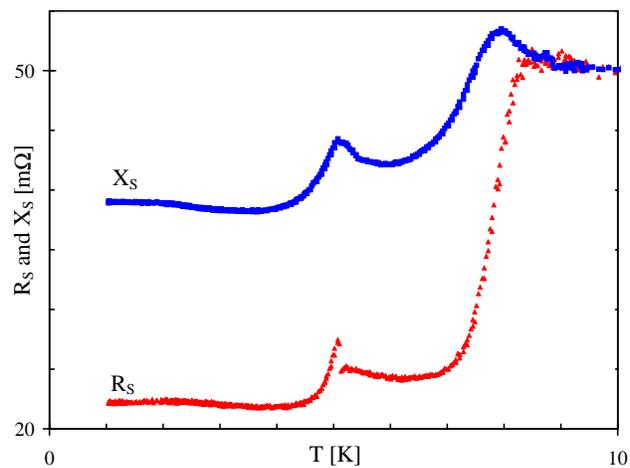} \caption{$R_s$ and
$X_s$ of HoNi$_2$B$_2$C. The AFM transition at $5.2\,$K is clearly
visible.}  \label{Fig4} \end{figure}

 A detailed plot of $R_s(T)$ vs. T in the region from 2K to 10K is
shown in Fig.~\ref{Fig4}. The drop in $R_s$ at the superconducting
transition temperature of 8K is clearly seen. In all conventional
superconductors, and in YNi$_2$B$_2$C as shown above, $R_s$ decreases
monotonically from the normal state value $R_n$. Remarkably, in
HoNi$_2$B$_2$C $R_s$ starts to increase again at around 6K. This is
due to pair-breaking which accompanies the development of the
antiferromagnetic state. At the AFM transition, $R_s$ shows a ${\em
peak}.$ Peaks in $R_s$ are never seen in conventional superconductors,
although non-monotonic dependence has been observed in
Y$_1$Ba$_2$Cu$_3$O$_{1-\delta}$
\cite{Bonn92}.

It is very interesting to study the reactance as a function of
temperature, also shown in Fig.~\ref{Fig4}. Below the superconducting
transition at $8K$, $X_s$ increases due to the buildup of superfluid,
as for the other superconductors, and then starts to decrease. However
the AFM transition intervenes and appears as a peak, due to
pairbreaking.

At low temperatures, both $R_s$ and $X_s$ are seen to {\em increase
with decreasing T}. Although this is similar to that observed in
TmNi$_2$B$_2$C, there is no AFM transition at a comparable
temperature. Thus our data indicates a possibly new source of magnetic
scattering below about $ 3.75K$.

 {\bf Complex Conductivity and Superfluid density}

\begin{figure} \epsfxsize=8.5cm \epsffile{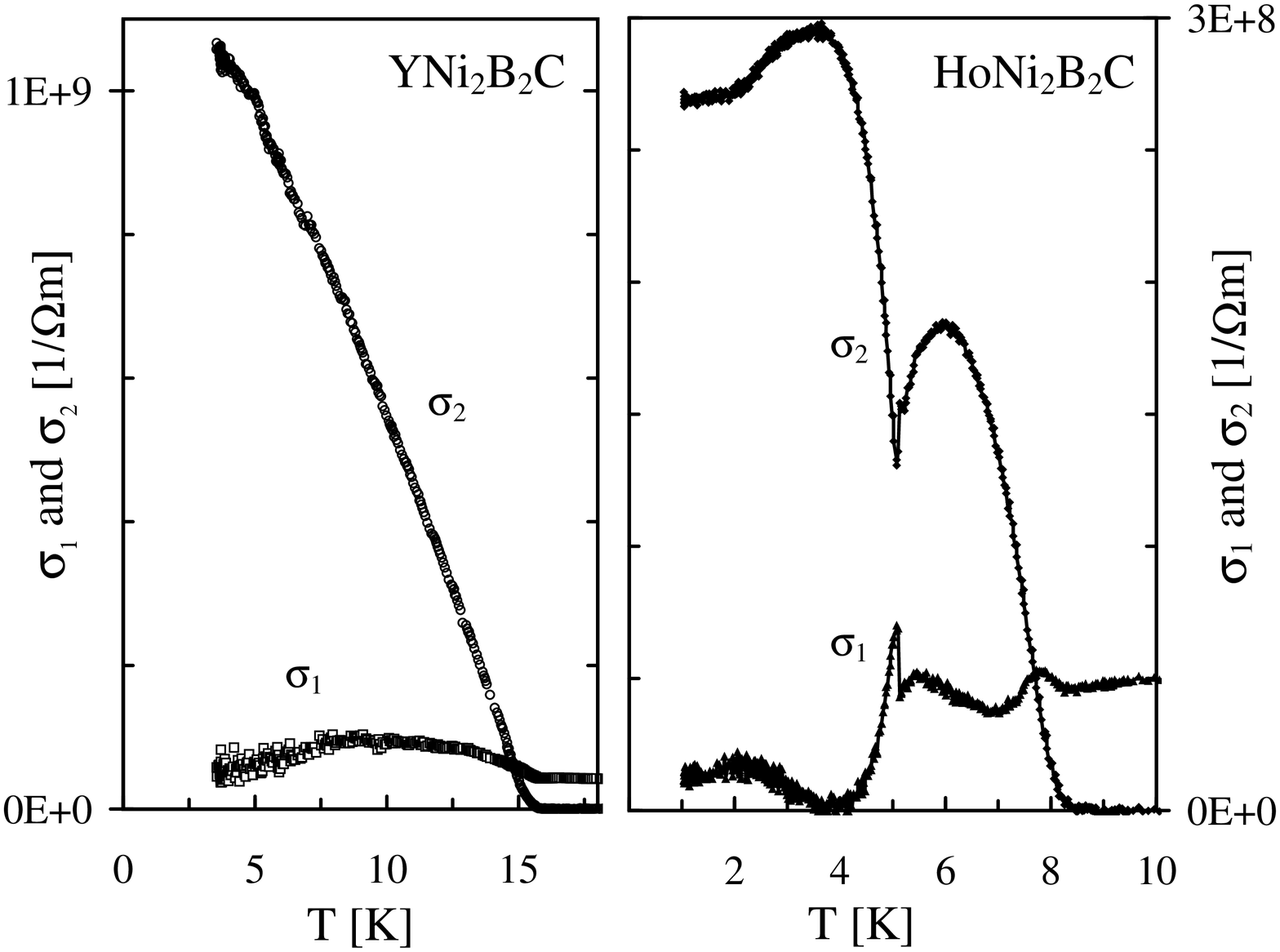} \caption{Real
($\sigma_1$) and imaginary ($\sigma_2$) parts of the complex
conductivity of (left) YNi$_2$B$_2$C and (right) unpolished
HoNi$_2$B$_2$C sample. $\sigma_2$ is proportional to the superfluid
pair density.}  \label{Fig5} \end{figure}

 The real and imaginary parts of the conductivity $\sigma _1$ and
$\sigma _2$ of YNi$_2$B$_2$C are shown in Fig.~\ref{Fig5}. Note that a
residual $R_{s,\,res}$ was subtracted while computing the complex
conductivity. $\sigma _2$ has a ``conventional'' temperature
dependence in that it rises smoothly at $T_c$ to a large value at low
temperatures. However the detailed temperature dependence does not fit
any conventional form such as $(1-t^4)$ of the 2-fluid model or a BCS
form. Instead $\sigma _2$ appears to be well described by a $(1-t)$
temperature dependence, except for a very slight curvature. $\sigma
_1$ rises from its normal state value and has a peak at around
$0.5T_c$. This is not the behavior expected from BCS coherence factors
since there the peak is around $0.9 T_c$. Instead this is similar to
the behavior of $\sigma _1$ of YBa$_2$Cu$_3$O$_{7-\delta}$ where there
is a peak at around $0.4T_c$ \cite{Bonn92}.

In HoNi$_2$B$_2$C, $\sigma _2$ and hence $n_s$ are found to initially
increase (see Fig.~\ref{Fig5}). However the AFM transition arrests
this increase, and {\em reduces the superfluid density}, although
superconductivity is not completely destroyed since $\sigma _2$ does
not become zero. Below the AFM transition, $\sigma _2$ recovers.  It
is interesting that the behavior of $\sigma _2$ very closely mirrors
that of $H_{c2}$ \cite{pcc8}. The AFM transition is also reflected in
the temperature dependence of $\sigma _1$.

 {\bf Discussion and Comparison with Theory}

\begin{figure}
\epsfxsize=8.5cm  \epsffile{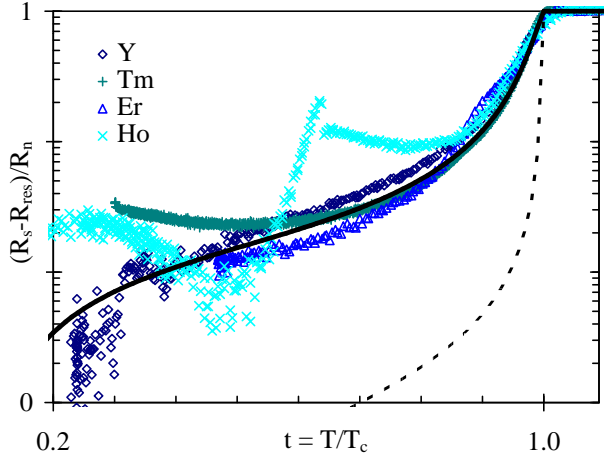}
\caption{Normalised surface resistance of LnNi$_2$B$_2$C \,
  (Ln = Y, Tm, Er, Ho) in the superconducting states vs. reduced
  temperature. The residual $R_s (T \rightarrow 0)$ has been subtracted
  out in each case.
  Also shown are BCS calculations with $\Delta (0)/\mathrm{k}T_c =1.74$
  (dashed line) and 0.45 (solid line).}  \label{Fig6} \end{figure}

 In order to address the issue of the nature of the superconducting
order parameter, we plot the normalized surface resistance $
(R_s(T)-R_{s,\,res})/R_s(T_c)$ vs. $t=T/T_c$ for all the superconductors
that were measured. The resulting plot shown in Fig.~\ref{Fig6} is
noteworthy in that the scaled plot appears to indicate a common,
underlying temperature dependence, independent of sample, except of
course for the expected deviations near $T_N$ in HoNi$_2$B$_2$C and
the low temperature rise for HoNi$_2$B$_2$C and TmNi$_2$B$_2$C. Also
shown in Fig.~\ref{Fig6} is a comparison with numerical calculations
based on the BCS theory using a Mattis-Bardeen complex conductivity in
the local limit \cite{Mattis58}. The gap parameter $\Delta (0)/kT_c$
was used as an input parameter. The data do not agree with the
numerical calculations if the mean-field value $\Delta (0)/kT_c=1.74$
is used. Instead a much lower value $0.49$ fits the data better. This
appears to be in conflict with measurements which suggest a BCS s-wave
state for the borocarbides. Indeed gap ratios of $1.45$ to $1.95$ for
YNi$_2$B$_2$C have been reported \cite{pcc1,TEkino94,THasegawa94}.

The data differ from the simple BCS analysis in three ways: (1) the
broader temperature dependence which can be modeled as a smaller gap
ratio, (2) the peaks at the AFM transition, and (3) the low
temperature rise in HoNi$_2$B$_2$C and TmNi$_2$B$_2$C. The latter two
features can be attributed to pair-breaking by the magnetic
constituents Ho and Tm. While the first feature could arise from
sample inhomogeneities at the surface, an interesting question is
whether the broadened transition could be due to an unidentified
source of pairbreaking which should then be present in all the
compounds. Equally noteworthy is the absence of any signature of the
AFM transition in the $Er$ superconductor. Neutron scattering studies
of HoNi$_2$B$_2$C \cite{pcc6} have shown the existence of a modulated
magnetic structure from $6\,$K to about $4.7\,$K. Similar features are
not observed in ErNi$_2$B$_2$C \cite{SKSinha95,JZarestky95}.
Pairbreaking is significant only when there exists magnetic fields
over length scales comparable to the coherence length, and this
appears to occur in HoNi$_2$B$_2$C, but not in ErNi$_2$B$_2$C.

In conclusion, the microwave properties of several borocarbide
superconductors reveal novel features of the superconducting
state. The superfluid density which is obtained from the data displays
striking features due to the influence of magnetic pairbreaking. The
data reveal interesting results on the interplay of magnetism and
superconductivity which deserve further experimental and theoretical
studies.

Work at Northeastern was supported by NSF-DMR-9223850. Ames Laboratory
is operated for the U. S. Department of Energy by Iowa State
University under Contract No. W-7405-Eng-82. Work at Ames was
supported by the Director for Energy Research, Office of Basic Energy
Sciences.

\narrowtext

\end{document}